# Two recipes for repelling hot water

Timothée Mouterde [1,2], Pierre Lecointre[1,2], Gaëlle Lehoucq[3], Antonio Checco[4], Christophe Clanet[1,2] & David Quéré [1,2]

Although a hydrophobic microtexture at a solid surface most often reflects rain owing to the presence of entrapped air within the texture, it is much more challenging to repel hot water. As it contacts a colder material, hot water generates condensation within the cavities at the solid surface, which eventually builds bridges between the substrate and the water, and thus destroys repellency. Here we show that both "small" (~100 nm) and "large" (~10 μm) model features do reflect hot drops at any drop temperature and in the whole range of explored impact velocities. Hence, we can define two structural recipes for repelling hot water: drops on nanometric features hardly stick owing to the miniaturization of water bridges, whereas kinetics of condensation in large features is too slow to connect the liquid to the solid at impact.

[1] Physique et Mécanique des Milieux Hétérogènes, UMR 7636 du CNRS, ESPCI, PSL Research University, 75005 Paris, France. [2] LadHyX, UMR 7646 du CNRS, École polytechnique, 91128 Palaiseau, France. [3] Thales Research and Technology, Route Départementale 128, 91767 Palaiseau, France. [4] Department of Mechanical Engineering, Light Engineering Lab, Stony Brook, New York 11794-2300, USA. These authors contribute equally: Timothée Mouterde, Pierre Lecointre. Correspondence and requests for materials should be addressed to T.M. (email: timothee.mouterde@polytechnique.org) or to D.Qéré. (email: david.quere@espci.fr)

When brought in contact with water, hydrophobic microfeatures on solids trap air, which lubricates the solid/liquid contact and renders water much more mobile than on conventional solids. As a consequence, most rough, hydrophobic materials repel rain in a dry atmosphere, resulting in spectacular rebounds after impact[1–3]: owing to its inertia, the impinging water first spreads as it would on plastic or glass, but the subjacent air then allows it to recoil and take off after a contact time of typically 10 ms. In contrast, humid conditions or dew repellency are much more challenging[4,5], as water nuclei condensing in the texture are then at the scale of the solid cavities, which fills the lubricating air layer and thus destroys superhydrophobicity. Many natural hierarchical surfaces, such as lotus leaves or artificial substrates covered by waxy microposts, indeed fail at repelling water in such conditions[6–8]. This limitation was recently circumvented by using nanopillars[9–11], whose scale can minimize the force of adhesion of water with the nuclei present in the nanocavities. This size effect can be amplified by making the nanopillars conical, a shape found to favour the jumping of condensing microdroplets as they coalesce[10,12–18].

In this context, it is not surprising that hydrophobic microtextures most often lose their superhydrophobicity when impacted by hot drops[19,20], except if the solid itself is hot[21]. The contact, even short, of hot water with a colder substrate promotes condensation within the microcavities at the solid surface, so as to bridge the incoming water to the substrate. How to repel hot water is an issue that has been poorly addressed despite its importance in industrial processes[22] such as clothing, coating, painting or windshield design[23]. This class of questions also includes the early stages of ice formation and accretion[24,25], potentially leading to serious damage to aircrafts, power lines, dams or wind turbines, when the typical time for phase change becomes shorter than the contact time of the impinging water. Hence, the idea to shorten the contact, which was achieved by tailoring large surface "defects"[26,27].

We wonder here how just the texture size may control the behaviour of hot water at impact and evidence two classes of texture able to robustly repel it. A first recipe consists in placing nanometric features (~100 nm) at the solid surface, which limits the size of water bridges and thus weakens the drop adhesion. A second recipe consists in having relatively "large" features (~10 μm), for which the construction of bridges is too slow to induce sticking during the brief contact at impact. We also explore the case of intermediate feature size, allowing us to test our model and to provide quantitative specifications for designing materials that can reflect hot water.

## Results

**Repellency failure.** Experiments are performed with model texture with size ranging from 100 nm to 10 μm. We use silica or silicon sculpted with cylindrical pillars (radius $a$, height $h$) disposed on a square lattice with pitch $p$ (Fig. 1a). All textures having a comparable geometry ($a \sim 0.1\,h$ and $p \sim h$), our materials are characterized by their pillar height, of respectively ~100 nm, ~1 μm and ~10 μm. The corresponding samples A, B and C, and their fabrication[10,28] are described in the Methods section. Pillars make surfaces rough and we classically define the roughness factor $r = 1 + 2\pi ah/p^2$ as the ratio of total to apparent surface areas. Texture is finally rendered hydrophobic by vapour deposition of 1H,1H,2H,2H-perfluorodecyltrichlorosilane and such treatment on flat silicon provides an advancing angle $\theta_a$ for water of 120° ± 2°, which jumps to $\theta_a = $ 167°, 168° and 169° (± 2°) on the rough materials A, B and C, respectively. The corresponding receding angles are $\theta_r = \theta_a - \Delta\theta = $ 140°, 143° and 152° (± 3°), providing contact angle hysteresis $\Delta\theta_A = $ 27°, $\Delta\theta_B = $ 25° and $\Delta\theta_C = $ 17°.

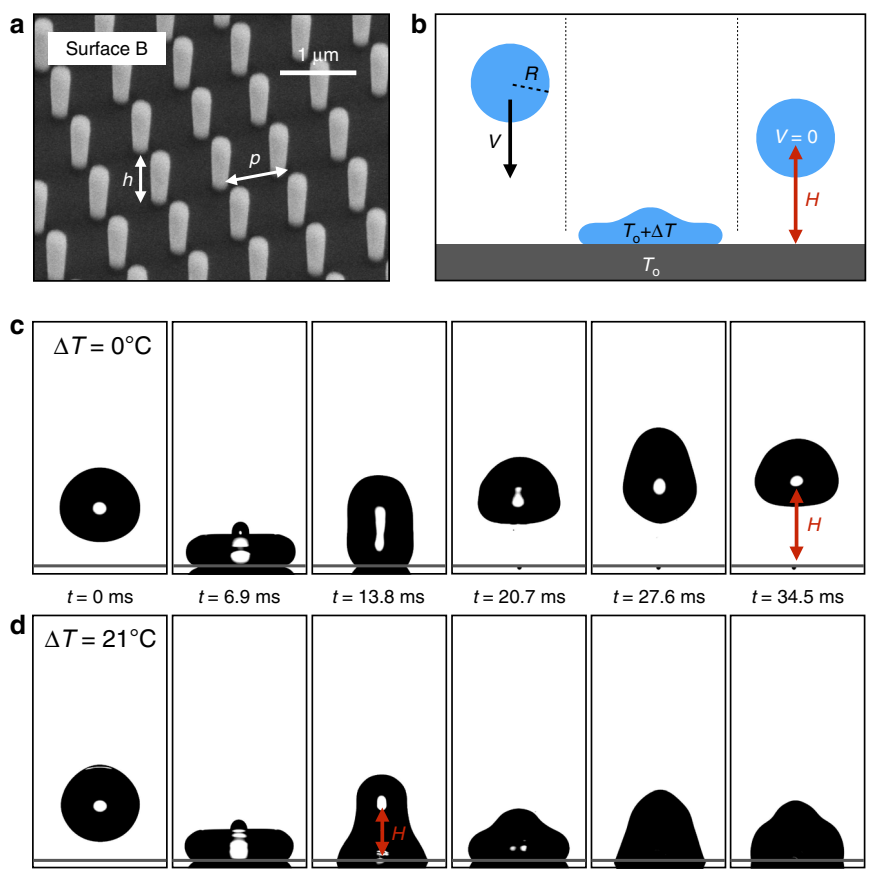

**Fig. 1** Impact of hot drops on textured solids. **a** Scanning-electron viewgraph of substrate B. It is covered by pillars with height $h = 900$ nm spaced by $p = 840$ nm. Scale bar, 1 μm. **b** Schematic of the experiment: a water drop with radius $R$ and temperature $T_o + \Delta T$ impacts at velocity $V$ a superhydrophobic substrate kept at $T_o$. We measure the height $H$ reached after impact, from which we deduce the restitution coefficient $\varepsilon$ of the shock. **c**, **d** Snapshots of water drops ($R = 1.4$ mm) impacting the substrate B at $V = 40$ cm s$^{-1}$. Images are separated by 6.9 ms and the temperature difference $\Delta T$ between the drop and the substrate is either 0 °C (**c**) or 21 °C (**d**). Water bounces in the first case and gets stuck in the second case

Our goal is to determine how water repellency is affected by condensation at impact, which we control through the temperature of impinging drops. Water is brought to a temperature $T_o + \Delta T$ and dispensed from a syringe kept at the same temperature, so as to form drops with a radius $R = 1.40 \pm 0.05$ mm (a second radius is tested in the Supplementary Information). Water does not cool down during its fall, the dispensing height $L \approx 1$ cm being such that the falling time $(2L/g)^{1/2} \approx 40$ ms is negligible compared with the thermalizing time $\rho C_p R/H_T \approx 40$ s, where $\rho = 1000$ kg m$^{-3}$ and $C_p = 4180$ J kg$^{-1}$ K$^{-1}$ are the density and thermal capacity of water, respectively, and $H_T \approx 100$ W m$^{-2}$ K$^{-1}$ is the heat transfer coefficient. Substrates kept at $T_o = 24 \pm 1$ °C and in a hygrometry of 32 ± 2% are impacted by drops impacting at a velocity $V = 40 \pm 5$ cm s$^{-1}$ (Fig. 1b) and we display in Fig. 1c,d high-speed snapshots of impacts on surface B ($h = 900$ nm).

Without temperature difference ($\Delta T = 0$ °C, Fig. 1c), water bounces off the solid, as also observed for all of our samples (Fig. 2a and Supplementary Movie 1). This regular superhydrophobic behaviour[1,2,29] can be understood by comparing the water dynamic pressure $\rho V^2$ and the Laplace pressure $\gamma/p$ opposing the penetration in pillars, where $\gamma$ is the water surface tension. This comparison entails a local Weber number $\rho V^2 p/\gamma$, a quantity always smaller than 0.01 in our experiments. The impacting liquid remains at the pillar tops, which makes rebounds possible and limits air compression within the pillars. The situation is quite different when elevating the temperature of water. As seen in Fig. 1d, an impinging drop brought to 45 °C ($\Delta T = 21$ °C) sticks to the sample B after impact, as revealed by the modest elevation $H$ of its centre of mass ($H \approx R$): the surface fails at repelling hot water.

**Texture repelling hot water.** However, hot water can be repelled by other texture. As seen in Fig. 2a–c and in Supplementary Movies 2 and 3, drops with $\Delta T = 21$ °C and $\Delta T = 40$ °C are reflected by materials A and C, on which rebounds are similar to that at $\Delta T = 0$ °C (Fig. 2a). We also report in the Fig. 2a–c the behaviour of a fourth sample, called A', where the texture is still nanometric yet larger than for A (210 nm instead of 88 nm, see details in the Methods section). Drops still bounce on this sample for $\Delta T = 21$ °C but cease to be repelled for $\Delta T = 40$ °C, confirming that repellency depends in a non-trivial way on the texture scale and water temperature, which we further explore in this study.

We quantify the ability of a solid to repel hot water by introducing the restitution coefficient $\varepsilon$ of the impacting drops, a quantity known for $\Delta T = 0$ °C to be fixed at a value $\varepsilon_0$ function of the impact velocity and contact angle hysteresis[30]. Before impact, the kinetic energy of a drop with mass $M = (4\pi/3)\rho R^3$ is $E_b = (1/2)MV^2$. When a drop bounces, its centre of gravity rises to a height

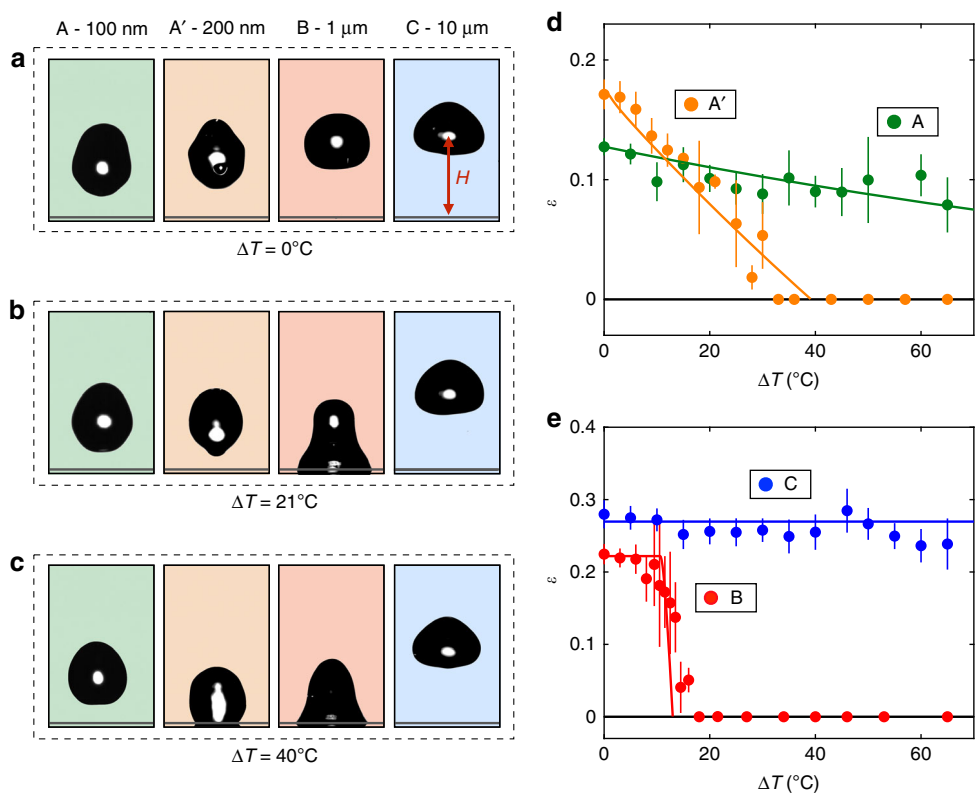

**Fig. 2** Bouncing behaviour of hot drops. **a–c** Impacting water drops ($R = 1.4$ mm and $V = 40$ cm s$^{-1}$) at their maximum bouncing height $H$ for $\Delta T = 0$°C (**a**), $\Delta T = 21$ °C (**b**) and $\Delta T = 40$ °C (**c**) on materials A, A', B and C with respective pillar heights of about 100 nm, 200 nm, 1 μm and 10 μm. **d**, **e** Coefficient of restitution $\varepsilon$ of the rebound as a function of the temperature difference $\Delta T$ between water and the solid surface ($R = 1.4$ mm, $V = 40$ cm s$^{-1}$). For the sake of clarity, we separate data on nanometric features (**d**) from data on micrometric features (**e**). Water bounces on samples A and C at all drop temperatures, while it gets trapped ($\varepsilon = 0$) on A' and B when $\Delta T$ exceeds ~40 °C and ~15 °C, respectively. Solid lines show equation (5). Error bars represent uncertainty of the measurement

$H > R$ and we express the potential energy after take-off as $E_a = Mg(H - R)$. The bouncing efficiency is then quantified by the coefficient $\varepsilon = E_a/E_b$, a quantity which is taken null when drops stick to the substrate ($H \approx R$). Our goal is to see how we deviate from the value $\varepsilon_0$ when water is hot, i.e., when condensation can take place during the impact.

We report the variation of $\varepsilon$ with $\Delta T$ in Fig. 2d,e for the surfaces A, A', B and C. We split the data in two graphs, to distinguish the behaviour on nano and microfeatures, which highlights the different nature of repellency in these two cases. Much information can be extracted from these plots: (1) At $\Delta T = 0$ °C (where condensation effects are marginal), the coefficient of restitution slightly varies with the texture (of slightly different contact angle hysteresis), with a typical value $\varepsilon_0 \approx 0.2$ characteristic of superhydrophobic rebounds at such impact velocity[30]. (2) On the smallest features (green data, $h \approx 100$ nm), drops systematically bounce. However, we observe that $\varepsilon$ slowly decreases with $\Delta T$, showing a small, continuous loss of kinetic energy at take-off as water gets warmer. This effect is amplified when using larger features (orange data, $h \approx 200$ nm), for which the decrease of $\varepsilon$ with $\Delta T$ becomes strong enough to intercept the abscissa axis in the range of explored temperature: drops hotter than 60 °C do not bounce on the substrate A'. (3) Observations are quite different with a micrometric texture. On the largest one (blue data, $h \approx 10$ μm), $\varepsilon$ is quasi-independent of $\Delta T$ ($\varepsilon = 0.27 \pm 0.04$) in the whole range of explored temperatures, $0 < \Delta T < 65$ °C. This is a surprising result, as we expect condensation to stick water all the more efficiently as $\Delta T$ increases. At smaller scale (red data, $h \approx 1$ μm), the behaviour is markedly different: after a small plateau, $\varepsilon$ tumbles around $\Delta T_c \approx 15$ °C and water sticks to the surface above this value, as already seen in Fig. 1c,d.

## Discussion

Contrasting with static situations where the smaller the texture, the better the water repellency in humid conditions[10], the fact that hot water bounces on 10 μm features reveals an original dynamical mechanism. When hot water contacts a colder superhydrophobic material, water nuclei form and grow within the texture (Fig. 3a). If they fill the elementary cells enclosed by neighbouring pillars, the resulting water bridges connect and stick the drop to its substrate (Fig. 3b). The formation of a bridge requires a time $\tau$ that can be evaluated. We assume that condensation is driven by a diffusive flux of water from the evaporating interface to the growing nucleus, whose respective vapour mass concentrations are $c_{sat}(T_o + \Delta T)$ and $c_{sat}(T_o)$. Denoting $\Delta c_{sat}(\Delta T) = c_{sat}(T_o + \Delta T) - c_{sat}(T_o)$, the diffusive flow rate scales as $D\Delta c_{sat}/h$, where $D \approx 20$ mm$^2$ s$^{-1}$ is the diffusion coefficient of vapour in air. Integrating this rate over the cell surface area $p^2$ and time $\tau$ gives the mass $\rho h p^2$ of the filled cell, which yields:

$$\tau \sim \rho h^2 / D\Delta c_{sat} \quad (1)$$

As it hits the solid and spreads along it at a velocity $V$, the impinging water draws vapour within and along the texture, which adds a convective term to the diffusive growth of the nucleus. The typical velocity $U$ of this vapour flow is deduced from the balance of viscous stresses at the liquid/vapour interface below the drop. Denoting $\eta$ and $\eta_v$ as the water and vapour viscosities, we simply write this balance as: $\eta V/R \sim \eta_v U/h$, where we neglect the friction of vapour around the pillars and thus slightly overestimate the convective flux. A vapour speed $U$ scaling as $(\eta h/\eta_v R) V$ is maximum for the tallest features ($h = 10$ μm), where it typically reaches 10 cm s$^{-1}$. Hence, the Péclet number $Pe = Uh/D$ comparing convective and diffusive flux is found to be at most 0.1 for $h = 10$ μm and much smaller for shorter features, which justifies our assumption of diffusive growth for the nucleus.

In usual conditions ($T_o \approx 24$ °C) and for $\Delta T \approx 10$ °C, we have $\Delta c_{sat} \approx 10$ g m$^{-3}$, which leads to $\tau \sim 1$ ms for $h \approx 1$ μm. $\tau$ increases by four orders of magnitude as $h$ rises from 100 nm to 10 μm, and it can be compared with the contact time $\tau_r$ of bouncing drops. $\tau_r$ being the response time of a spring with mass $\rho R^3$ and stiffness $\gamma$, we have[31]: $\tau_r \sim (\rho R^3/\gamma)^{1/2}$, whose weak dependency on $\rho$ and $\gamma$ allows us to neglect its variation with temperature. The time $\tau_r$ is on the order of 10 ms for millimetric drops and thus possibly comparable to $\tau$: There is a texture height, in the range of a few micrometres, for which we expect the two times to be equal, which eventually allows us to model the different impacts.

When the condensation time $\tau$ is larger than the bouncing time $\tau_r$, water nuclei are smaller than the roughness height $h$ (Fig. 3c) and thus do not connect the impacting drop to the substrate: $\varepsilon$ is constant and equal to $\varepsilon_0$, its value at $\Delta T = 0$ °C:

$$\varepsilon = \varepsilon_0 \quad (2)$$

This mechanism explains the observations for the material C in Fig. 2d, a case where the large height of the pillars implies $\tau > \tau_r$. Hence, a material with a tall texture can dynamically repel hot

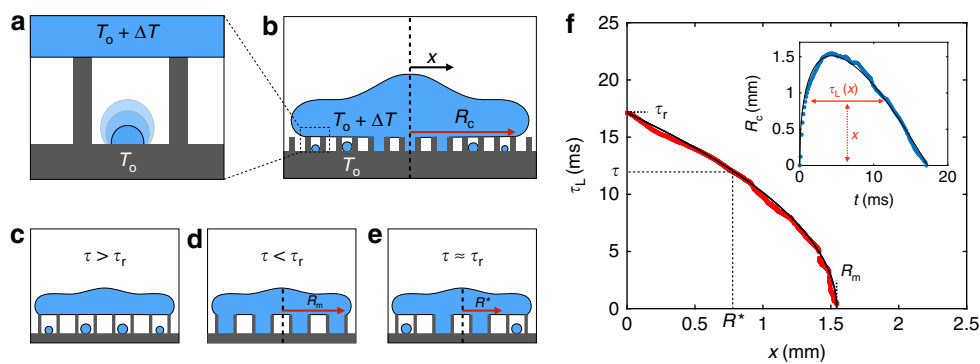

**Fig. 3** Dynamics of adhesion. **a** Elementary cell containing a growing condensation nucleus fed by the hot water above. **b** Sketch of the spreading drop. $x$ is the radial distance from the impact point and condensation nuclei are larger at small $x$, due to a higher local contact time $\tau_L$. **c**–**e** Pinning mechanism as a function of $\tau$ and $\tau_r$, the condensation and bouncing times. In the first case (**c**, $\tau > \tau_r$), there is no bridge between the substrate and the drop; in the second case (**d**, $\tau \ll \tau_r$), bridges form with a probability $N$ below the drop; in the third case (**e**, $\tau \approx \tau_r$), bridges (with the same probability $N$) only connect the drop on a radius $R^* < R_m$, the maximum contact radius at impact. **f** Local contact time $\tau_L$ as a function of the distance $x$ for a drop bouncing on surface B ($R = 1.4$ mm and $V = 40$ cm s$^{-1}$). The solid line shows equation (4), with $\tau_r = 17.4$ ms and $R_m = 1.55$ mm. For $\tau_L > \tau$, the drop is bridged to the surface on a disk with radius $x = R^*(\tau)$. Insert: contact radius $R_c$ as a function of time for the same experiment. The solid fit $R_c(t) \approx 2(RVt)^{1/2} - (3\gamma/2\rho R)^{1/2} We^{1/4} t$ is discussed in the text

water, whereas it would lose its superhydrophobic character if exposed longer to humidity.

The opposite limit ($\tau < \tau_r$) concerns short pillars, when cavities enclosing a condensation droplet instantaneously fill at impact (Fig. 3d). As for hot water sitting on a cold texture[10], the adhesion force is $F \approx 4\pi R_m \gamma N$, where $R_m$ is the maximum contact radius (sketched in Fig. 3d) and $N$ is the probability of having a water nucleus in a cell. If we denote $n$ as the number of nuclei per unit area, $N$ is just $\min(nrp^2, 1)$, where the increase of $n$ with $\Delta T$ ($n \approx 0.06\, \Delta T\, \mu m^{-2}$) can be determined by static measurements[10]. The energy $E_{adh}$ induced by condensation and dissipated during bouncing is given by the work of $F$ on the radius $R_m$, which yields $E_{adh} = 2\pi\gamma R_m^2 \min(nrp^2, 1)$. Hence, we deduce the coefficient $\varepsilon = \varepsilon_0 - E_{adh}/E_b$:

$$\varepsilon = \varepsilon_0 - 4\pi\gamma R_m^2 \min(nrp^2, 1)/MV^2 \quad (3)$$

At small nucleus density ($n(\Delta T) < 1/rp^2$), $n$ linearly varies with $\Delta T$ and equation (3) predicts a linear decrease of $\varepsilon$ with temperature, as seen with sample A. For a small texture, we have $nrp^2 < 1$, which yields: $\varepsilon = \varepsilon_0 - 4\pi\gamma R_m^2 rp^2 n/MV^2$. Drawn with a green line in Fig. 2d for the experimental value $R_m = 1.55$ mm, this law is found to quantitatively fit the data, explaining the persistence of bouncing at any water temperature and the slight decay of $\varepsilon$ with $\Delta T$. Three parameters in the model, namely $\gamma$, $\rho$ and $n$, depend on temperature but the weak variations of $\gamma$ and $\rho$ with $T$ make negligible their influence on the fit. The model also explains the behaviour observed with the sample A' (orange data), whose texture is twice larger than A. Then, we expect a stronger decrease of the function $\varepsilon(\Delta T)$ (sensitive to the quantity $rp^2$) and even strong enough to intercept the axis $\varepsilon = 0$. Hence, we quantitatively understand the transition to sticking for the orange data in Fig. 2d and more generally what is the feature size limit for repelling hot water with a nanotexture (see Supplementary Information for details).

The final case concerns the transition regime where both timescales have comparable magnitude ($\tau \approx \tau_r$). As sketched in Fig. 3e, condensation is favoured close to the impact point: water spends there more time than at the drop periphery, where it comes later and recedes earlier. We can introduce a local contact time $\tau_L(x)$, denoting $x$ as the distance from the drop centre (Fig. 3b). By definition, $\tau_L$ is the contact time $\tau_r$ at the impact point ($\tau_L(0) = \tau_r$) and it vanishes at the drop periphery ($\tau_L(R_m) = 0$). If we denote $R^*$ as the distance where the local contact time $\tau_L$ and the condensation time $\tau$ are equal, $\tau_L(R^*) = \tau$, we can distinguish two zones (Fig. 3e): for $x > R^*$, condensation is too slow to connect the drop to the solid and this area does not contribute to adhesion; for $x < R^*$, condensation bridges the material to the drop on a disk with radius $R^*$ and the adhesion energy $E_{adh}$ is determined by replacing in equation (3) $R_m$ by $R^*$, the radius of the adhesive area.

As seen in Fig. 3f, the function $\tau_L(x)$ can be deduced from the time evolution of the contact radius $R_c$ (insert in the figure). It can be also modelled by assuming that the contact dynamics can be divided in two phases: (1) at small time, the drop "sinks" at velocity $V$ in the solid with the shape of a truncated sphere[32–34], which provides a Hertzian scaling: $R_c(t) \sim (RVt)^{1/2}$. This relationship was found to hold all along the spreading[32], with a numerical factor of ~2. (2) The recoiling drop[35] is a pancake with thickness $z$ (Fig. 3a) that dewets at the Taylor–Culick velocity $(2\gamma/\rho z)^{1/2}$. Considering that the average height $z$ of this pancake is given by a balance between inertia and surface tension[36], i.e., $z \approx (4R/3)We^{-1/2}$ (with $We = \rho V^2 R/\gamma$), we eventually get in this regime $R_c(t) \sim -(3\gamma/2\rho R)^{1/2} We^{1/4} t$. We finally assume that both contributions are additive, which yields: $R_c(t) \approx 2(RVt)^{1/2} - (3\gamma/2\rho R)^{1/2} We^{1/4} t$, a function drawn with a solid line in the insert of Fig. 3f where it nicely fits the data at all times. This description is valid at modest Weber number $We$, when no rim forms[37], similar to here where we have $We \approx 3$. The non-monotonic character of $R_c(t)$ implies that the equation $R_c = x$ has two solutions in time for $0 < x < R_m$, which we denote as $t_1$ and $t_2$ ($t_1 < t_2$). By definition, we have $\tau_L(x) = t_2 - t_1$, a quantity that can be extracted analytically from the expression of $R_c(t)$. We find:

$$\tau_L(x) \approx a(1 - x/b)^{1/2} \quad (4)$$

where $a = (8/3)\,(\rho R^3/\gamma)^{1/2} \approx \tau_r$ and $b = RWe^{1/4} \approx R_m$ (see the Supplementary Information for details). Equation (4) is drawn in Fig. 3f and observed to fit the data for $a \approx 17.4$ ms and $b \approx 1.6$ mm, i.e., the experimental values of $\tau_r$ and $R_m$ that themselves nicely compare with the expected ones, $a \approx 15.9$ ms and $b \approx 1.8$ mm. From the explicit expression $\tau_L(x) \approx \tau_r\,(1 - x/R_m)^{1/2}$, we deduce the adhesion radius and find $R^* = R_m\,[1 - (\tau/\tau_r)^2]$ for $\tau \leq \tau_r$, and $R^* = 0$ for $\tau > \tau_r$. Combining these equations with equation (3) in the transition regime, $\varepsilon = \varepsilon_0 - 4\pi\gamma R^{*2} \min(nrp^2, 1)/MV^2$, we get a general expression for the coefficient of restitution $\varepsilon$:

$$\varepsilon = \varepsilon_0 - 4\pi\gamma R_m^2 \max^2([1 - (\tau/\tau_r)^2], 0) \min(nrp^2, 1)/MV^2 \quad (5)$$

In order to compare this prediction with our data, we estimate $\Delta c_{sat}$ using Rankine formula (see details in the Methods section). The only adjustable parameter is the numerical factor $\alpha$ in equation (1), $\tau = \alpha \rho h^2/D\Delta c_{sat}$. Drawn with solid lines in Fig. 2d,e with $\alpha = 8$, equation (5) is found to describe the whole ensemble of data. We recover the two limit cases, $\tau > \tau_r$, where condensation does not affect bouncing (equation (2), sample C and sample B at small $\Delta T$), and $\tau < \tau_r$, where condensation is instantaneous (equation (3), samples A and A'), a regime sensitive to temperature. Moreover, equation (5) predicts the failure of repellency at intermediate pillar height (sample B). The transition is indeed found to be abrupt and to occur at the temperature given by the equality $\tau(\Delta T_c) = \tau_r$. Drops then cease to bounce, which we indicate by the line $\varepsilon = 0$. It is noteworthy that our test of equation (5) for another drop radius confirms its ability to describe all the regimes (Supplementary Fig. 1). We also studied the influence of the impact velocity (up to 1 m s$^{-1}$) and ambient hygrometry, and found again that the model is robust enough to capture the ensemble of data (Supplementary Figs 2, 3 and 4).

The predictive character of equation (5) eventually allows us to construct a phase diagram based on the temperature difference $\Delta T$ and pillar height $h$, fixing the other parameters (homothetic samples with $p = h$, $a = h/6$, $r \approx 2$, $\varepsilon_0 = 0.2$, $R = 1.4$ mm and $V = 40$ cm s$^{-1}$, all values comparable to that in our experiments). In the resulting phase diagram (Fig. 4), green and red colours distinguish bouncing from sticking. As found experimentally, the sticking region is indeed observed at intermediate pillar height and for $\Delta T > \Delta T_c$, where $\Delta T_c$ is given by $\varepsilon(\Delta T_c) = 0$ in equation (5) (solid line in the figure). Comparison with experiments can be refined by marking whether drops bounce or stick using green or red symbols. Experiments are performed with the samples A, A', B and C, to which we add data obtained with a fifth surface B' where pillar characteristics are $a = 100$ nm, $h = 600$ nm and $p = 560$ nm ($r \approx 2.2$). For samples A and C only (extreme values of $h$, $h = 88$ nm and $h = 10\,\mu m$), we remain in the bouncing regime whatever the water temperature, whereas bouncing/sticking transitions are observed at intermediate texture size (samples A', B' and B). The location of the transition is in good agreement with the prediction for $\Delta T_c$, confirming for instance the non-monotonic character of $\Delta T_c$ with the pillar height. The two extreme values $h_1$ and $h_2$ below and above which rebounds are observed at all $\Delta T$ (dashes in the figure) can be expressed explicitly from equation (5), as shown in the Supplementary Discussion.

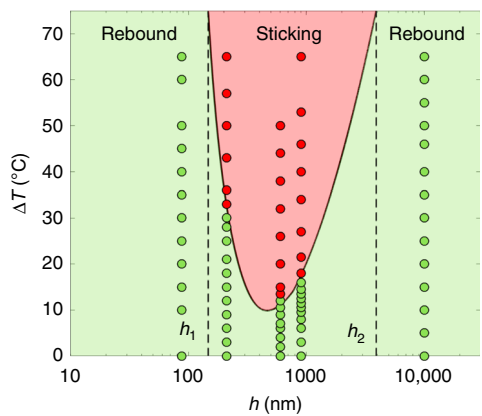

**Fig. 4** Phase diagram for hot water impacting hydrophobic texture. The two control parameters are the water/substrate temperature difference $\Delta T$ and the pillar height $h$. The critical temperature $\Delta T_c$ defining the frontier between bouncing (green region) and sticking (red region) is calculated by solving $\varepsilon = 0$ in equation (5), with $p = h$, $a = h/6$, $r \approx 2$, $\varepsilon_0 = 0.2$, $R = 1.4$ mm and $V = 40$ cm s$^{-1}$, all values comparable to that in our experiments. The lower and upper bounds $h_1$ and $h_2$ for which bouncing is observed whatever the drop temperature are stressed with vertical dashes. Observations (green and red symbols for bouncing and sticking, respectively) are made for the five tested samples (A, $h = 88$ nm; A', $h = 210$ nm; B', $h = 600$ nm; B, $h = 900$ nm; C, $h = 10$ μm) and found to be in good agreement with the model

Our study shows the existence of two structural recipes for repelling hot water on a hydrophobic, textured material. (1) Tall features dynamically prevent the texture filling and the formation of water bridges between impacting drops and the solid material. (2) Small features miniaturize the bridges and thus their sticking abilities. In both cases, repellency is robust, as it is observed in the whole range of explored impact velocity and water temperature. The existence of two scenarii of repellency is reflected by differences in the repellency itself: although rebounds are found to be nearly unsensitive to water temperature in case 1, warmer drops are repelled slightly slower in case 2, due to the multiplication of water bridges when the impacting water is hotter. It would be interesting to mix the two kinds of texture to see what is the dominant scenario in such a case. On a more fundamental note, playing with the texture size gives access to the dynamics of condensation at submicrometric scales, a measure known to be particularly challenging. Our findings imply to consider the feature scale when developing new design able to reduce the bouncing time[26–28] for anti-icing or anti-fogging properties. However, the scale is not the only geometrical parameter: for instance, modifying (at constant height) the distance between the features or their order (square, hexagonal, etc.) should be alternative ways to design materials that repel hot water. It would finally be worth exploring what happens when water impacts repellent materials at room temperature in a rarefied atmosphere. Both the evaporation rate[38] and impact characteristics[39] are dramatically affected in a low-pressure environment, which might modify water repellency[40]. Non-condensable gases in air act as a neutral medium in our experiment. Replacing them by pure vapour should favour water condensation, whereas an increased rate of evaporation can conversely induce self-taking off (trampolining) of water, i.e., an increase of repellency[38]. On the whole, the interplay of impact at low pressure with the texture geometry and scale should be a subject of interest for the future.

## Methods

**Surface A**. This surface is fabricated by combining block-copolymer self-assembly with anisotropic plasma etching in silicon, which provides large-area (cm$^2$) textures with ~10 nm feature size and long-range order. Posts have a radius $a = 15$ nm and a height $h = 88$ nm, and they are disposed on a rhombus network with side $p = 52$ nm. The roughness factor $r$ is $r_A \approx 4.5$, and the water advancing and receding angles are $\theta_a = 167 \pm 2°$ and $\theta_r = 140 \pm 2°$, respectively.

**Surface A'**. The texture is a square lattice of pillars fabricated by electron-beam lithography and anisotropic plasma etching in silica. The pillar radius, height and spacing are respectively $a = 35$ nm, $h = 210$ nm and $p = 140$ nm. The pillar density and aspect ratio are ~20% and $h/2a = 3$, respectively, and the roughness factor is $r_{A'} \approx 3.4$. The water advancing and receding angles are $\theta_a = 155 \pm 3°$ and $\theta_r = 132 \pm 3°$, respectively.

**Surface B**. The texture is a square lattice of pillars fabricated by electron-beam lithography and anisotropic plasma etching in silica. The pillar size, height and spacing are respectively $a = 150$ nm, $h = 900$ nm and $p = 840$ nm. The pillar density and aspect ratio are ~10% and $h/2a = 3$, respectively, and the roughness factor is $r_B \approx 2.2$. The water advancing and receding angles are $\theta_a = 168 \pm 2°$ and $\theta_r = 143 \pm 2°$, respectively. Surface B' (used in Fig. 4) is made the same way, with $a = 100$ nm, $h = 600$ nm and $p = 560$ nm.

**Surface C**. This surface fabricated by photolithography and deep reactive ion etching is a square lattice of pillars. The pillar size, height and spacing are respectively $a = 1.25$ μm, $h = 10$ μm and $p = 10$ μm. The pillar density and aspect ratio are ~5% and $h/2a = 4$, respectively, and the roughness factor is $r_B \approx 1.8$. The water advancing and receding angles are $\theta_a = 169 \pm 2°$ and $\theta_r = 152 \pm 2°$, respectively.

**Thermodynamic quantities**. The water vapour concentration $c_{sat}(T)$ at temperature $T$ is given by Dalton's law: $c_{sat}(T) = \rho_{sat}(M_w/M_{air})(P_{sat}(T)/P_0)$, where $M_w$ and $M_{air}$ are the respective molar masses of water and air, $P_{sat}(T)$ is the saturated vapour pressure of water at temperature $T$ and $P_0$ is the atmospheric pressure. $P_{sat}(T)$ is given by the empirical Rankine formula: $P_{sat}(T) = P_0 \exp(13.7 - 5120/T)$. With this relation, we can estimate $c_{sat}$ at temperatures $T_0$ and $T_0 + \Delta T$, a useful information in equation (1).

### Data availability
The data that support the plots within this paper and other findings of this study are available in the main text and in the Supplementary Information. Additional information is available from the authors upon reasonable request.

### Acknowledgements
We thank Direction Générale de l'Armement (DGA) for contributing to the financial support, Rose-Marie Sauvage for her constant interest, Mathilde Reyssat, Stéphane Xavier, Atikur Rahman and Charles Black for their help in microfabrication, and Thales for cofunding this project. P.L. thanks the Ecole polytechnique for the financial support (Monge Fellowship).

### Author contributions
T.M. conceived the project. T.M., P.L., C.C. and D.Q. designed the project. G.L. and A.C. provided samples and discussed the project. T.M. and P.L. performed experiments and analyses. T.M., P.L., C.C. and D.Q. built the model. T.M., P.L. and D.Q. wrote the manuscript with inputs from all other authors.

### Additional information
**Supplementary Information** accompanies this paper at https://doi.org/10.1038/s41467-019-09456-8.

**Competing interests:** The authors declare no competing interests.

**Reprints and permission** information is available online at http://npg.nature.com/reprintsandpermissions/

**Journal peer review information**: *Nature Communications* thanks Sanjeev Chandra and the other anonymous reviewers for their contribution to the peer review of this work. Peer reviewer reports are available.

**Publisher's note:** Springer Nature remains neutral with regard to jurisdictional claims in published maps and institutional affiliations.